\documentclass[pra,aps,superscriptaddress,10pt,floatfix,amsmath,amssymb,showkeys]{revtex4-2}
\usepackage{graphicx} 
\usepackage{dcolumn}
\usepackage{bm}
\usepackage{physics}
\usepackage{lipsum}
\usepackage{orcidlink}

\graphicspath{{images/}}

\begin{document}
\title{Mesoscopic Quantum Communication via Photon-Number Moments}
\author{Gabriele Cenedese\,\orcidlink{0000-0003-4531-1112}}
\affiliation{Instituto de Física Interdisciplinar y Sistemas Complejos (IFISC), UIB–CSIC
UIB Campus, E-07122 Palma de Mallorca, Spain.}
\author{Alex Pozzoli\,\orcidlink{0009-0003-8917-7725}}
\affiliation{Como Lake Institute of Photonics, Department of Science and High Technology, University of Insubria, Via Valleggio 11, I-22100 Como, Italy}
\author{Luca Razzoli\,\orcidlink{0000-0002-9129-2154}}
\affiliation{Department of Physics ``Alessandro Volta'', University of Pavia, Via Bassi 6, I-27100 Pavia, Italy; INFN, Sezione di Pavia, Via Bassi 6, I-27100 Pavia, Italy}
\author{Alessia Allevi\,\orcidlink{0000-0002-1972-3124}}
\email{alessia.allevi@uninsubria.it}
\affiliation{Como Lake Institute of Photonics, Department of Science and High Technology, University of Insubria, Via Valleggio 11, I-22100 Como, Italy}
\begin{abstract}
Mesoscopic optical states are a promising resource for quantum communication, combining robustness against losses with the preservation of genuine quantum features. Here, we propose a quantum communication protocol in which information is encoded in the first and second moments of the photon-number distributions of classical optical states, and then decoded by photon-number-resolving detectors. Security relies on the nonclassical photon-number correlations of a twin-beam state transmitted alongside the signal in the quantum channel, providing an experimentally accessible security witness against both intercept-resend and beam-splitter attacks investigated in this work. 
Numerical simulations performed in experimentally accessible parameter regimes support the feasibility and security of the proposed communication protocol, yielding nonzero key generation rates under the considered eavesdropping attacks, and motivating its future experimental implementation.
\end{abstract}
\maketitle
\section{Introduction}\label{sec_intro}
Quantum key distribution (QKD) aims at generating and distributing secret keys between two or more parties by exploiting the rules of quantum mechanics \cite{cariolaro} and its characteristic traits, namely quantum superposition and entanglement \cite{benenti,diamanti}.
The technological framework underpinning single-photon QKD is nowadays well established and relies on well-known QKD protocols \cite{gisin}, such as BB84 \cite{bennett1984quantum,bennett1992experimental} and E91 \cite{ekert1991}, together with single-photon sources \cite{eisaman2011revsciinst}, optimal receivers---including avalanche photodiodes (APDs) and, more recently, superconducting nanowire single-photon detectors (SNSPDs) \cite{flamini,cozzolino,boaron}---and comprehensive security analyses \cite{scarani2009RevModPhys,xu2020RevModPhys}.
Although the technology has reached a high level of maturity, with several components and even complete QKD systems now commercially available, single-photon QKD is intrinsically fragile, as it relies on single-photon-level signals that are inherently vulnerable to losses and experimental imperfections \cite{dimario2018robust,dimario2,li2020enhanced}. In addition, most single-photon sources are probabilistic or heralded, and are therefore characterized by a low repetition rate, which limits how quickly information can be transmitted \cite{azuma2024heralded}.

The mesoscopic intensity regime \cite{arimondo}---i.e., the intermediate level of photon-number states between single-photon and classical light levels---represents an appealing operating domain, offering enhanced robustness against losses while still preserving quantum properties of the optical states. Being comparatively underexplored, this regime still lacks well-established secure communication protocols, as already discussed in Ref.~\cite{razzoli25}. Addressing this gap requires both suitable detection schemes and the extension of eavesdropping attacks.
Receivers in single-photon QKD typically rely on standard on/off detection schemes based on APDs and SNSPDs. Beyond the single-photon regime, however, such detection schemes become overly restrictive compared to the richer photon-number statistics involved. In the mesoscopic intensity regime, photon-number-resolving (PNR) detectors \cite{becerra,oe24} represent a more suitable class of detectors for quantum communication protocols, as they enable proper reconstruction of the photon-number distribution of the received optical states \cite{chesi19,cassina,endo}. Communication protocols operating in this regime must therefore be designed to fully exploit the PNR capability of these detectors \cite{cattaneo2018,notarnicola2023,notarnicola2025}.
This also calls for a dedicated security analysis, revisiting conventional eavesdropping attacks in  the mesoscopic intensity regime. Specifically, for protocols that fundamentally rely on PNR detectors, i.e., on the capability of counting the number of photons, the typical eavesdropping strategies include the intercept-resend (IR) \cite{usenko} and the beam-splitting (BS) attacks \cite{calsamiglia}, where the sent signals are partially intercepted and modified by Eve. Detecting such attacks requires the development of robust criteria based on the available non-classical resources and the specification of appropriate acceptance thresholds.\\
In this work, we propose a communication protocol based on multi-mode twin-beam (TWB) states generated by spontaneous parametric down-conversion and receivers endowed with PNR capabilities. TWB states, which exhibit photon-number entanglement \cite{pla22}, constitute the nonclassical resource that ensures the security of the protocol. In fact, applying the nonclassicality criterion based on the noise reduction factor $R$ between the two arms of the TWB allows us to detect the presence of eavesdropper's attacks, and thus to interrupt the communication \cite{scirep22}. However, unlike standard communication protocols such as E91, where information is encoded in the entangled states, in our protocol information is encoded in suitable optical states superimposed on one arm of the TWB \cite{razzoli25}. This allows one to select states with specific statistical features and to fully leverage the PNR capabilities of the receiver by encoding information not only in the mean photon number but also in higher moments of the photon-number distribution.
More specifically, we consider four super-Poissonian optical states with different variances and two distinct mean values, resulting in an alphabet of eight symbols.
Compared with binary encoding, this choice increases the channel capacity, although it makes state discrimination more challenging. To address this task, we investigate state discrimination using different machine-learning classifiers, comparing their performance in terms of classification accuracy \cite{kotsiantis,osisanwo}.     
Furthermore, the robustness of the protocol against IR and BS attacks is assessed as a function of the fraction of data intercepted by Eve and the BS reflectivity, respectively. 
Upon optimization of the parameters characterizing the light sources and the transmission channel, the resulting communication scheme proves robust against both considered attacks, even when information is encoded in a limited number of data.
This robustness is quantified in terms of the true positive rate (TPR) in a symbol-resolved security test based on the noise reduction factor, which complements the discrimination of the symbols in terms of the statistical properties of detected states.

Finally, we show that this security check is also correlated to the key generation rate (KGR), which is evaluated from the mutual information (MI) shared between the sender and the receiver in the presence of an eavesdropper.
Overall, these findings pave the way for a more comprehensive security characterization of communication protocols operating in the mesoscopic intensity regime and provide a foundation for their future experimental implementation.\\ 
The paper is organized as follows: First, in Sec. \ref{sec_methods} we introduce the moment-based eight-symbol encoding and the noise reduction factor ($R$) witness for the TWB. We then characterize the discrimination performance of the receiver and analyze the protocol under IR and BS attacks in Sec. \ref{sec_simulations}. In Sec. \ref{sec_discussion} we compare the $R$-based TPR with the MI-based KGR, showing that the two quantities are strongly correlated for the considered attack models. Finally, in Sec. \ref{sec_conclusions} we draw our conclusions.
\section{Theoretical framework}\label{sec_methods}
We encode information in the first two moments of photon-number distributions. We consider four classes of classical optical states, each characterized by a distinct photon-number distribution. For each distribution, two mean photon numbers are employed. Since the four distributions exhibit different photon-number variances for a fixed mean, each symbol is identified by the pair consisting of its mean photon number and variance.
The resulting eight-symbol alphabet can be resolved using the PNR capabilities of the receiver.

\begin{figure*}[!t]
\centering
\hfill
\includegraphics[width=0.9\textwidth]{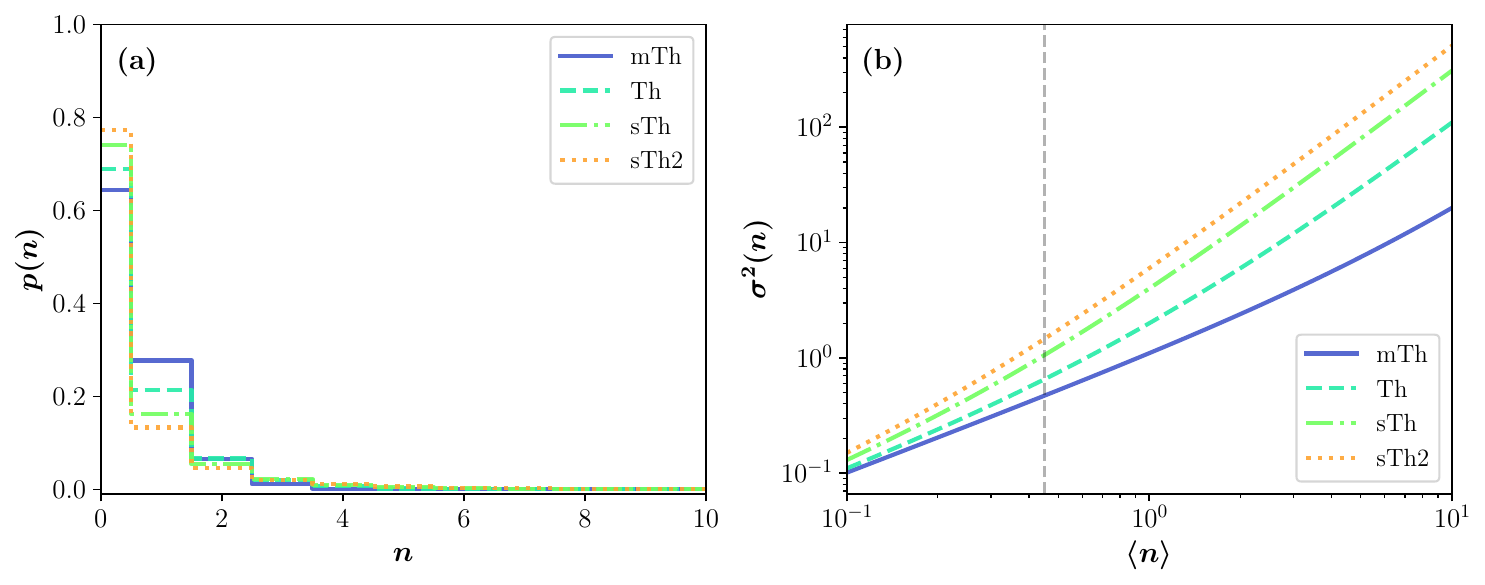}
\caption{(a) Photon number distributions of the multi-mode pseudo-thermal state (mTh), single-mode pseudo-thermal state (Th), super-thermal
state from a speckled-speckle field (sTh), and second-harmonic of single-mode pseudo-thermal state (sTh2), all of them having a mean value $\langle n \rangle = 0.45$, $\mu_s=1$ for Th, STh and STh2, while $\mu_s=10$ for mTh, and $\mu_{s2}=1$ for STh. (b) Variance as a function of the mean value for the same states as in panel (a); the vertical dashed line highligths $\langle n \rangle = 0.45$.}
\label{distributions}
\end{figure*}

We focus on classical states with a super-Poissonian distribution, as they have been shown to directly affect measurable nonclassicality criteria \cite{oe21}. More specifically, we consider pseudo-thermal states obtained by passing a laser beam through a diffuser as well as super-thermal states obtained by manipulating either in a linear \cite{bianciardi} or nonlinear way \cite{ol15} the light exiting the diffuser.
Under the assumption of $\mu_s$ equally-populated states, the multi-mode pseudo-thermal distribution reads as \cite{mandel}
\begin{equation} \label{multiTh}
    p_{\rm mTh}(n) = \frac{(n+ \mu_s -1)!}{n!(\mu_s-1)!(\langle n \rangle/\mu_s + 1)^{\mu_s}(\mu_s/\langle n \rangle +1)^n},
\end{equation}
where $\langle n \rangle$ is the mean number of photons. In the case of $\mu_s =1$, Eq.~(\ref{multiTh}) describes the statistics of a single-mode pseudo-thermal state
\begin{equation} \label{1Th}
    p_{\rm Th}(n) = \frac{\langle n \rangle^n}{(1+\langle n \rangle)^{(n+1)}}.
\end{equation}
If the $\mu_s$ modes of the aforementioned distribution are further scattered, a speckled-speckle field is obtained \cite{goodman2007}. The resulting distribution is super-thermal and also characterized by one more parameter, that is the number of modes $\mu_{s2}$ produced by the second diffuser. The corresponding photon-number distribution reads as \cite{pla23}
\begin{eqnarray} \label{superTh}
    p_{\rm sTh}(n) &=& \frac{1}{\Gamma[\mu_s] \Gamma[\mu_{s2}]n!} \left( \frac{\mu_s \mu_{s2}}{\langle n \rangle} \right)^{(\mu_s+ \mu_{s2}- |\mu_s- \mu_{s2}|)/2}\\
    & \times & \Gamma \left[ \frac{2n+\mu_s+ \mu_{s2}- |\mu_s- \mu_{s2}|}{2} \right] \Gamma \left[ \frac{2n+\mu_s+ \mu_{s2}+ |\mu_s- \mu_{s2}|}{2} \right] \nonumber\\
    & \times & U \left[ \frac{2n+\mu_s+ \mu_{s2}- |\mu_s- \mu_{s2}|}{2}, 1- |\mu_s- \mu_{s2}|, \frac{\mu_s \mu_{s2}}{\langle n \rangle} \right], \nonumber
\end{eqnarray}
where $\Gamma[j]$ is the Gamma function and $U(a,b,c)$ is the confluent hypergeometric function of the second kind. In the case $\mu_s = \mu_{s2} =1$, Eq.~(\ref{superTh}) reduces to
\begin{eqnarray}\label{superth1mode}
p_{\rm sTh}(n) &=& \frac{1}{\langle n \rangle}\Gamma\left[1+ n\right] U\left[(1+n),1, \frac{1}{\langle n \rangle}\right],
\end{eqnarray}
On the other hand, if the distribution in Eq.~(\ref{multiTh}) is up-converted in a second-order nonlinear crystal, the second-harmonic of the incident field is obtained, whose photon-number distribution reads as \cite{ol15}
\begin{eqnarray}\label{superthermSH}
p_{\rm sTh2}(n) &=& \frac{\Gamma[1/2 + n +\mu_s/2] \Gamma[n + \mu_s/2]}{4 \sqrt{\pi} n! (\mu_s - 1)! \{ \langle n \rangle / [\mu_s (1+ \mu_s)] \}^{(\mu_s + 1)/2} }\nonumber\\
&\times& U \left [\frac{1}{2} + n + \frac{\mu_s}{2}, \frac{3}{2}, \frac{\mu_s (\mu_s + 1)}{4 \langle n \rangle} \right ]
\end{eqnarray}
under the assumption that all the $\mu_s$ speckles are frequency doubled with the same efficiency and properly selected. In the case of $\mu_s = 1$, the expression simplifies to
\begin{eqnarray}\label{superthermSH1mode}
p_{\rm sTh2}(n) &=& \frac{1}{2 \sqrt{\pi} \langle n \rangle}\Gamma[1/2 + n] U \left [1 + n, \frac{3}{2}, \frac{1}{2 \langle n \rangle} \right ].
\end{eqnarray}
All the aforementioned photon-number distributions (shown in Fig.~\ref{distributions}(a) for the case $\langle n \rangle = 0.45$), associated with classical states, have a variance that can be expressed as \cite{mandel}
\begin{equation} \label{variance}
\sigma^2(n) = \langle n \rangle (a \langle n \rangle + 1),
\end{equation}
where $a$ is a distribution-dependent coefficient. For the classical optical states considered for encoding the alphabet, Eq.~\eqref{variance} reads as follows: 
\begin{itemize}
\item for the single-mode pseudo-thermal state in Eq.~(\ref{1Th}), in which $a=1$,
\begin{equation} \label{var_th}
\sigma_{\rm Th}^2(n) = \langle n \rangle \left( \langle n \rangle + 1 \right);
\end{equation}
\item for the multi-mode pseudo-thermal state in Eq.~(\ref{multiTh}), in which $a = 1/\mu_s$,
\begin{equation} \label{var_mth}
\sigma_{\rm mTh}^2(n) = \langle n \rangle \left( \frac{\langle n \rangle}{\mu_s} + 1 \right);
\end{equation}
\item for the photon-number distribution in Eq.~(\ref{superth1mode}), in which $a = 3$,
\begin{equation} \label{var_sth}
\sigma_{\rm sTh}^2(n) = \langle n \rangle \left(3 \langle n \rangle + 1 \right);
\end{equation}
\item and finally, for the photon-number distribution in Eq.~(\ref{superthermSH1mode}), in which $a = 5$,
\begin{equation} \label{var_shth}
\sigma_{\rm sTh2}^2(n) = \langle n \rangle \left(5 \langle n \rangle + 1 \right).
\end{equation}
\end{itemize}
As evident from Eqs.~\eqref{var_th}–-\eqref{var_shth}, the difficulty of discriminating among the eight possible states ultimately depends on the mean photon number, $\langle n \rangle$. In particular, increasing $\langle n \rangle$ improves the discrimination performance as the corresponding variances are more separated \cite{izumi,becerra,becerra1,muller}, as shown in Fig.~\ref{distributions}(b).
Remarkably, Eqs. \eqref{var_th}--\eqref{var_shth} are invariant under Bernoullian detection, thus remaining formally the same except for the use of the number of detected photons, $m$, instead of the number of incident photons, $n$.\\ 
To enhance attack-detection capabilities, the security of the communication protocol relies on the use of nonclassical resources in the communication channel. 
Since the receiver employs PNR detection, we consider multi-mode TWB states as the relevant nonclassical resource, owing to their photon-number entanglement.
By assuming that the $\mu$ spatio-spectral modes \cite{machulka2014} characterizing the TWB are equally populated, the generated state can be written as the tensor product of $\mu$ identical TWB states \cite{epl10,silberhorn2016},
\begin{equation} \label{multiTWB}
\vert \Psi^{\mu}_{\rm TWB} \rangle = 
\bigotimes_{k=1}^\mu \sqrt{1- \lambda^2}
\sum_{\nu=0}^{\infty} \lambda^{\nu} \vert \nu \rangle_k \otimes \vert \nu\rangle_k,
\end{equation}  
where $k$ labels the modes, $\nu$ is the photon number in each mode, $\lambda$ is defined through $\lambda^2 = \langle n \rangle / (\mu + \langle n \rangle)$, and $\langle n \rangle$ is the mean total number of photons in either of the two arms of the TWB.
The noise reduction factor provides a suitable criterion to prove the nonclassicality of such states, because it can be written in terms of measurable quantities and can include the contribution of additional signals superimposed on one or both arms of the TWB. 
In terms of incident photons, the noise reduction factor is defined as 
\begin{equation} \label{Rdetphot}
R = \frac{\sigma^2(n_1-n_2)}{\langle n_1 \rangle + \langle n_2 \rangle},
\end{equation}
where $\sigma^2(n_1-n_2)$ is the variance of the distribution of the photon-number difference between the two parties of the TWB and $(\langle n_1 \rangle + \langle n_2 \rangle)$ is the shot-noise level. Values $R<1$ prove that the states are nonclassically correlated, and this constitutes a sufficient condition for the entanglement \cite{agliati}.
When an additional signal is superimposed on one arm of the TWB state, both the variance and $R$ increase depending on the underlying statistical distribution.
Focusing on measurable quantities as in a real setting, their expressions in terms of the number of detected photons read, respectively,
\begin{equation} \label{noisyvariance}
\sigma^2(m)
= \sigma^2_{\rm TWB} + \sigma^2_{\rm s}
= \langle m \rangle \left( \frac{\langle m \rangle}{\mu} + 1 \right) + \langle m_s \rangle \left(a \langle m_s \rangle +1\right),
\end{equation}
and
\begin{equation} \label{noisyR}
R = 1- \frac{2\eta \langle m \rangle}{2 \langle m \rangle + \langle m_{\rm s} \rangle} + \frac{a \langle m_{\rm s} \rangle^2}{2\langle m \rangle + \langle m_{\rm s} \rangle},
\end{equation}
where $\langle m \rangle$ is the mean number of detected photons in one arm of the TWB, 
$\langle m_{\rm s} \rangle$ is the mean number of detected photons of the superimposed signal, $\eta$ is the global quantum efficiency of the detection system, and $a$ is the distribution-dependent coefficient introduced in Eq.~\eqref{variance}. The superimposed signal is uncorrelated with the TWB. 
\begin{figure*}[!t]
\centering
\hfill
\includegraphics[width=0.9\textwidth]{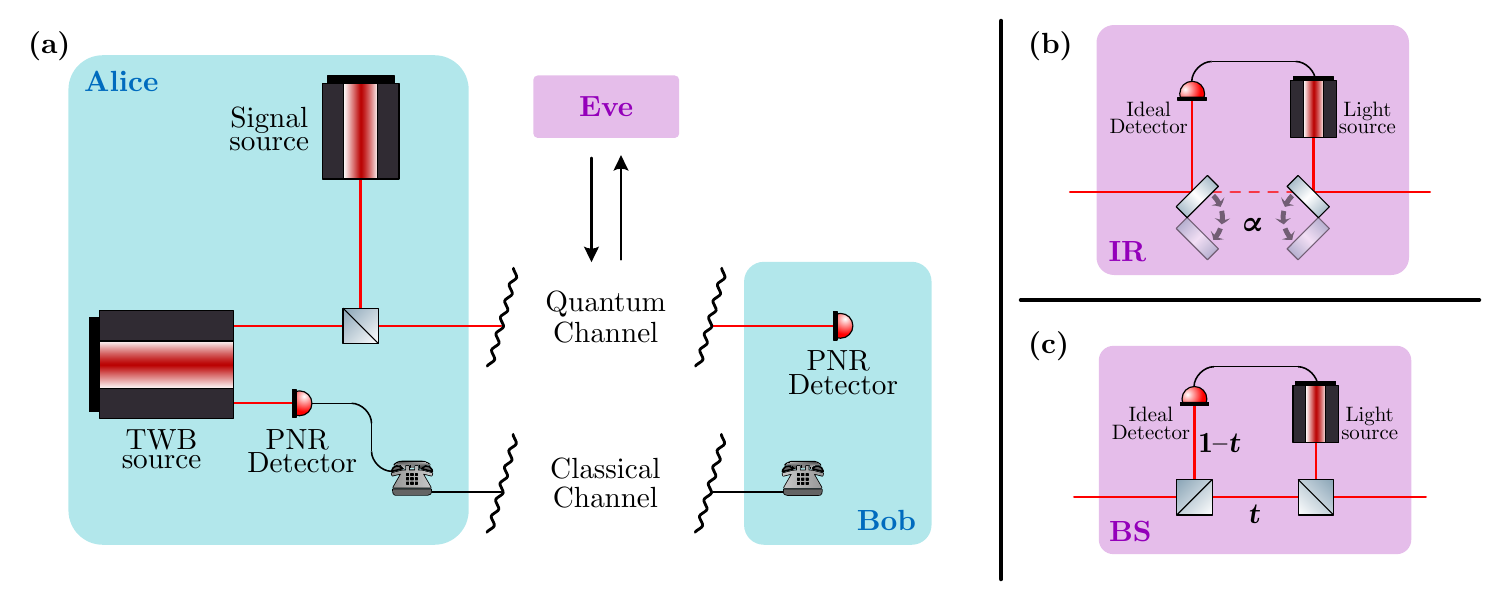}
\caption{(a) Sketch of the proposed communication scheme, where Alice encodes information into classical optical states of light with different photon-number distributions and sends them to Bob through a quantum channel together with the portion of a TWB state. Sketch of (b) the IR attack and (c) the BS attack performed by Eve. In (b) $\alpha$ is the fraction of data corresponding to a given symbol intercepted and replaced by Eve, while in (c) $t$ is the transmissivity of the BS and $(1-t)$ its reflectivity.}
\label{scheme}
\end{figure*}
\subsection{The communication protocol and the eavesdropping attacks} \label{sec:protocol}
In the protocol, Alice generates both the classical states in which she encodes information and the TWB states. As shown in the sketch of Fig.~\ref{scheme}(a), she sends one of the classical states superimposed on one arm of the TWB to Bob through the communication channel, while retaining the other arm. 
Encoding information in the first two moments of the photon-number distribution, and subsequently discriminating the transmitted state, requires for each symbol a data sample of sufficient length---i.e., a sufficiently high number of laser pulses---to reliably estimate statistical moments.
Both Alice and Bob are equipped with identical receivers based on PNR detectors, which allow them to obtain statistical information about the measured states.
The proper discrimination of the states depends on the kind of distribution, but also on the mean value of light. Indeed, the lower the mean value of the signal state, the harder the discrimination. This can be noticed both considering Eqs.~(\ref{noisyvariance}) and (\ref{noisyR}), where the dependence on the mean value of the signal is explicitly stated. 
In the following, we focus on a \textit{local-moment discrimination strategy} that relies on the mean and variance of the state measured by Bob (i.e., one arm of the TWB with the superimposed signal). In addition, we explore a \textit{correlation-assisted discrimination strategy} that uses the mean value of the state measured by Bob together with the noise reduction factor between the two arms of the TWB. In fact, it is fully determined by first- and second-order photon-number moments [see Eq. \eqref{Rdetphot}], and thus remains consistent with our moment-based discrimination framework. The only implication is that evaluating $R$ requires classical communication between Bob and Alice, as shown in Fig. \ref{scheme}(a). 

An eavesdropper, Eve, may exploit the large number of pulses used to encode each symbol to devise attacks aimed at extracting information.
Hereafter, we assume that Eve is also equipped with the same type of receiver as Alice and Bob, but with unit quantum efficiency, and consider two possible eavesdropping strategies: the IR and the BS attacks. In the IR attack [Fig.~\ref{scheme}(b)], Eve \textit{intercepts} a portion of light containing the signal used to encode a specific symbol, and she \textit{resends} a different light mimicking the stolen one, that is a light signal with the same mean value and variance as those of the photon-number distribution she measures. In this case, since each symbol is decoded from a finite data sample, where each entry records the number of detected photons in a laser pulse, security is investigated as a function of the number of data subtracted and replaced. 
In the BS attack [Fig.~\ref{scheme}(c)], Eve uses a BS with variable transmissivity $t$ to split the light sent to Bob, thereby accessing the entire data sample associated with a given symbol, rather than only a subset of pulses as in the IR attack. This operation reduces the mean value measured by Bob, making the attack easily detectable through a PNR detector\cite{gaidash2016}. To investigate a more challenging BS attack, we assume that Eve superimposes a new signal onto the beam transmitted by the BS, with the same mean value and variance as those of the distribution she measures. In this case, the security is investigated as a function of the reflectivity $(1-t)$. We determine the values of $(1-t)$ that allow Eve to discriminate the sent state, and the minimum reflectivity for which Bob can detect eavesdropping. In principle, in both attacks Eve can approach the discrimination of the signal Alice has sent to Bob not only via the local-moment discrimination strategy, but also via the correlation-assisted discrimination strategy, due to the vulnerability of the classical communication channel.
\subsection{Security criterion and discrimination performance metrics}
The nonclassical resource on which the security of the proposed communication protocol relies is represented by the nonclassical photon-number correlations of the TWB. Operating the communication protocol in a regime where the superimposed signal still preserves $R<1$, any eavesdropping attack that is uncorrelated with the TWB and leads to  $R>1$ can be readily detected. However, as we show in the following, this criterion is insufficient. In fact, depending on Eve's capabilities and available resources, effective attacks can be devised that still preserve $R<1$, while allowing Eve to correctly discriminate the signals. Therefore, security must be enforced by introducing more tailored acceptance criteria on the values of the noise reduction factor.
The capability of Bob (and Eve) in discriminating the sent signals is investigated for both local-moment and correlation-assisted discrimination strategies. In fact, although the noise reduction factor is highly sensitive to correlation-breaking perturbations introduced by Eve, its ability to discriminate the symbols depends critically on how the superimposed signal modifies the photon-number statistics. In contrast, the local-moment discrimination strategy directly probes the marginal statistics of Bob's arm and may therefore be more robust when the signal is primarily local. 
The discrimination performance can be quantified by the accuracy parameter 
\begin{equation}
{\rm ACC} = d_{\rm dis}/d_{\rm string},
\label{eq:accuracy}
\end{equation}
that is the ratio between the number of symbols correctly discriminated (by Bob or Eve), $d_{\rm dis}$, to the total size of the string, $d_{\rm string}$, transmitted by Alice. In the following, ${\rm ACC}_{\sigma^2(R)}$ denotes the accuracy of the local-moment (correlation-assisted) discrimination strategy.
\section{Simulations}\label{sec_simulations}
\subsection{The setting}
This work provides a proof-of-principle demonstration of the proposed quantum communication protocol through numerical simulations. The simulations are used to assess its feasibility, discrimination capability, and robustness against eavesdropping, while identifying suitable setup parameters---such as the mean photon numbers and the number of modes---under realistic experimental conditions, thus paving the way for future implementation.

We consider a multi-mode TWB with a mean value, measured by Alice, in the range $3.0 \leq \langle m_{\rm{Alice}} \rangle \leq 10.0$ and a number of modes $\mu = 100$. For the signals encoding information, we consider two possible mean values $\langle m_{s,{\rm L}} \rangle = 0.40$ and $\langle m_{s,{\rm H}} \rangle = 0.45$, where H (L) denotes high (low) mean value, that are significantly smaller than the mean value of the TWB.
This is due to the fact that, as discussed in  Ref.~\cite{oe21}, the mean value of the signal that can be superimposed on a portion of TWB while keeping the noise reduction factor below 1 depends on the mean value of TWB, the number of modes of the signal, the quantum efficiency of the detection chain, and the loss $\Delta$ affecting the transmission channel.
In the following, we consider loss in the range $0.5 \leq \Delta \leq 0.9$, and assume that it only affects the TWB (in practical implementations, this can be achieved by increasing the intensity of the classical signal superimposed onto the TWB arm to compensate for channel losses).

The eight-symbol alphabet is encoded using single-mode pseudo-thermal and single-mode super-thermal states, whose variances are reported in Eqs.~\eqref{var_th}, \eqref{var_sth}, and \eqref{var_shth}, respectively, and multi-mode pseudo-thermal states in Eq.~\eqref{multiTh} with a number of modes equal to $\mu_s = 10$, thereby ensuring a marked difference in the value of the variance (see Eq.~\eqref{var_mth}) with respect to the analogous single-mode case $\mu_s = 1$ in Eq. \eqref{1Th}.

Concerning the receiver, we assume that Bob and Alice are equipped with a PNR detector with quantum efficiency $\eta = 0.4$, which is the case of Silicon photomultipliers (SiPMs) operated in the visible spectral range \cite{S13360}. We also assume that the PNR capability of the receiver is sufficiently high (up to 50 photons\cite{arxiv2026}).
For each condition of the signal superimposed on the TWB (i.e., combination of mean value and photon-number distribution), 
we numerically generate a dataset of $10^5$ simulated laser pulses, where each entry represents the measured photon number.
This allows us to reliably estimate mean value, variance, and noise reduction factor for each superimposed signal.
We then generate symbols by applying a bootstrap resampling procedure \cite{efron,bipm} to the corresponding datasets, using samples of $d_{\rm sample}$ data points per symbol. 
The reference string adopted to compute the accuracy in Eq. \eqref{eq:accuracy} is constructed via bootstrap by generating 100 realizations of each of the eight symbols, resulting in a string of length $d_{\rm string} = 800$. The measurement procedure and the introduction of the loss in the arm of TWB sent to Bob have been performed according to a Bernoullian distribution. 
\subsection{Preliminary characterization}
\begin{figure*}[!t]
\centering
\hfill
\includegraphics[width=0.9\textwidth]{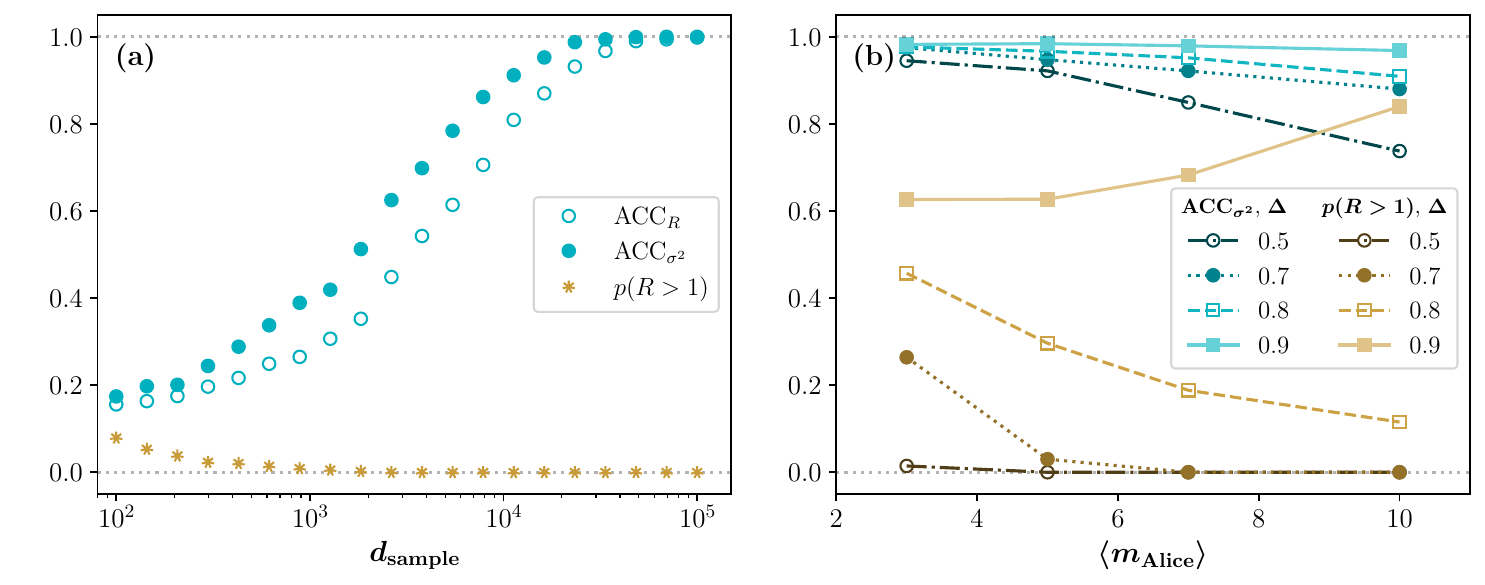}
\caption{
(a) Bob's discrimination accuracies ${\rm ACC}_{\sigma^2}$, ${\rm ACC}_R$, and $p(R>1)$ as a function of the data sample size $d_{\rm sample}$ per symbol, for loss $\Delta = 0.5$ and mean photon number of the TWB $\langle m_{\rm{Alice}} \rangle = 5$. (b) Bob's ${\rm ACC}_{\sigma^2}$ and $p(R>1)$ as a function of $\langle m_{\rm{Alice}} \rangle$, for $d_{\rm sample} = 2 \times 10^4$ and varying $\Delta$.
In both panels, blue tones refer to ${\rm ACC}$, while brown tones refer to $p(R>1)$. The superimposed signals have mean values $\langle m_{\rm s,L} \rangle = 0.40$ and $\langle m_{\rm s,H} \rangle = 0.45$.
}
\label{accuracyVSdataset}
\end{figure*}

The first and essential step is to determine the minimum sample size required to reliably discriminate the transmitted symbols in the absence of eavesdropping, assuming an intermediate loss $\Delta = 0.5$ in the quantum channel. Accordingly, we investigate the accuracies ${\rm ACC}_{\sigma^2}$ and ${\rm ACC}_{R}$ of the local-moment and correlation-assisted discrimination strategy, respectively, as a function of the number of data $d_{\rm sample}$ used to encode each symbol. For this preliminary study, discrimination is performed by a k-Nearest Neighbors (kNN) classifier. In addition, since the sample size also affects the estimate of the noise reduction factor, we investigate the probability of obtaining $R>1$, despite the superimposed signals being designed to preserve $R<1$. As shown in Fig.~\ref{accuracyVSdataset}(a), ${\rm ACC}_{\sigma^2}$ converges to 1 faster than ${\rm ACC}_{R}$, saturating to ${\rm ACC_{\sigma^2}} = 1$ for $d_{\rm sample} \gtrsim 2\times 10^4$, while the probability that $R>1$ decreases to 0 as the sample size increases. 
Based on these results, we conclude that the local-moment discrimination strategy outperforms the correlation-assisted one and therefore adopt it for symbol discrimination throughout the remainder of this work. Accordingly, we consider $d_{\rm sample}=2 \times 10^4$ as the minimum sample size to encode symbols. Although the noise reduction factor is not employed for discrimination, it will complement the local-moment discrimination strategy from a security perspective, allowing for eavesdropping detection. Indeed, its evaluation requires classical communication between Alice and Bob, a potential vulnerability, and the sacrifice of part of the transmitted data. While these additional requirements would negatively affect the discrimination performance, they are acceptable in the context of security analysis, where the noise reduction factor is used as an auxiliary criterion, as discussed in the following section.

Next, we address the role of losses affecting the transmission channel and, consequently, the TWB state.
As shown in Fig. \ref{accuracyVSdataset}(b), the accuracy ${\rm ACC}_{\sigma^2}$ increases with the amount of loss $\Delta$\, while it decreases as the mean value $\langle m_{\rm Alice} \rangle$ of the TWB increases. These behaviors can be understood in the light of the variance in Eq. \eqref{noisyvariance}: differences in the variance are more pronounced, i.e., the symbols are more distinguishable, when the mean value of the superimposed signal $\langle m_{\rm s} \rangle \sim 0.4-0.45$ is comparable to, or on the same order of, the mean value of the TWB, $\langle m_{\rm Alice} \rangle$.
Reversely, we observe a decrease of the accuracy at increasing values of $\langle m_{\rm Alice} \rangle$; this effect is particularly evident for $\Delta = 0.5$ (the minimum loss value considered), as in this case $\langle m_s \rangle$ remains always significantly lower than that of TWB.
In our setting, discrimination becomes therefore easier for less populated TWB states, as the superimposed signals have low mean value.
On the other hand, the probability of measuring $R>1$ increases for increasing $\Delta$, meaning that larger loss is detrimental for detecting nonclassical correlations. Also, $p(R>1)$ decreases as $\langle m_{\rm Alice} \rangle$ increases, as the superimposed signal becomes negligible compared to baseline set by the TWB, thus better preserving the nonclassical correlations. An exception to the latter statement is observed for $\Delta = 0.9$, for which the imbalance between the two arms of the TWB is so high that increasing $\langle m_{\rm Alice} \rangle$ yields an increase in $p(R>1)$, therefore a different strategy should be used in high loss scenarios.
Instead, by considering the different slopes of the curves corresponding to $\Delta<0.9$, we observe that the effect of increasing the loss $\Delta$ can be effectively mitigated by increasing $\langle m_{\rm Alice} \rangle$, i.e., by employing a more intense TWB. Conversely, for smaller values of $\Delta$, lower values of $\langle m_{\rm Alice} \rangle$ are required to observe $R<1$.
Combining the results obtained for ACC$_{\sigma^2}$ and $p(R>1)$, we identify the parameter set $\Delta = 0.5$ and $\langle m_{\rm Alice} \rangle = 5$ as the most representative operating condition for further investigation. This choice corresponds to a mean number of photons incident on Alice's detector equal to $\langle n \rangle = \langle m_{\rm Alice} \rangle/ \eta = 12.5$.

\begin{figure*}[!t]
\centering
\hfill
\includegraphics[width=0.9\textwidth]{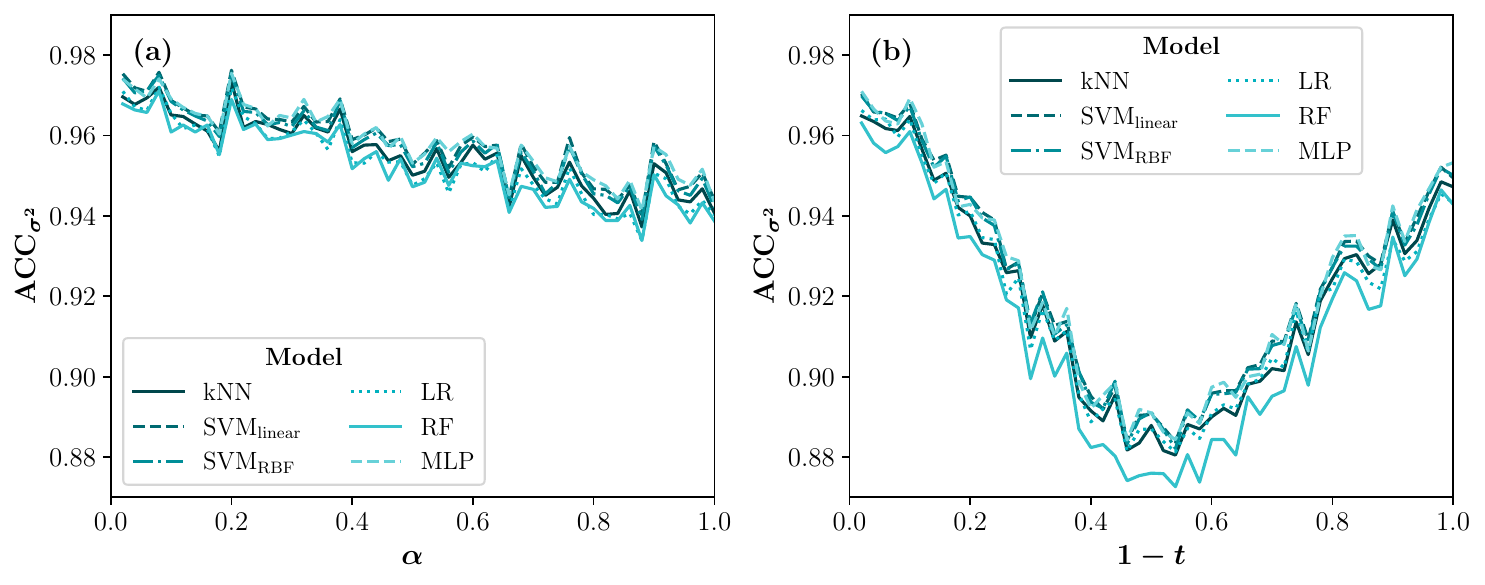}
\caption{(a) Bob's discrimination accuracy ${\rm ACC_{\sigma^2}}$ as a function of the fraction of intercepted data $\alpha$ in the IR attack and (b) as a function of the BS reflectivity $(1-t)$ in the BS attack. Different colors and line styles correspond to different classifiers: k-Nearest Neighbors (kNN), Linear Support Vector Machine (SVM$_{\rm linear}$), Support Vector Machine with Radial Basis Function kernel (SVM$_{\rm RBF}$), Logistic Regression (LR), Random Forest (RF), and Multi-Layer Perceptron (MLP). In all panels the parameters are $\langle m_{\rm s,L} \rangle = 0.40$, $\langle m_{\rm s,H} \rangle = 0.45$, $\langle m_{\rm{Alice}} \rangle = 5$, $\Delta = 0.5$, and $d_{\rm sample} = 2\times 10^4$.}
\label{IRattackML}
\end{figure*}

To characterize the behavior of the protocol under eavesdropping attacks, we consider two scenarios, namely the IR and BS attacks. Bob discriminates the signals according to the local-moment strategy, and we investigate the dependence of the results on the relevant attack parameters: the fraction of intercepted data per symbol, $\alpha = d_{\rm int}/d_{\rm sample}$, where $d_{\rm int}$ denotes the total number of intercepted data, for the IR attack, and Eve's beam-splitter reflectivity $(1-t)$ for the BS attack.
For the discrimination task, we consider six classical supervised machine-learning classifiers: kNN, Linear Support Vector Machine, Support Vector Machine with Radial Basis Function kernel, Logistic Regression, Random Forest, and Multi-Layer Perceptron. The goal is to compare their discrimination performance and 
investigate their ability to capture possible eavesdropping-induced signatures.
Figures~\ref{IRattackML}(a) and (b) show the accuracy ${\rm ACC}_{\sigma^2}$ as a function of $\alpha$ and $(1-t)$ for each classifier, respectively. We first observe that, in this setting, the different classifiers yield qualitatively similar results: all of them exhibit an overall decreasing trend in panel (a) and a minimum at $(1-t) = 0.5$ in panel (b). In particular the "U" shape of the accuracy in panel (b) is related to the term $t(1-t) \langle  m \rangle$, modeling the BS introduced by Eve, which modifies the expression of the variance measured by Bob.
Remarkably, ${\rm ACC}_{\sigma^2}(\alpha=0)={\rm ACC}_{\sigma^2}(1-t=0)$, since both cases correspond to the absence of eavesdropping. Similarly, ${\rm ACC}_{\sigma^2}(\alpha=1)={\rm ACC}_{\sigma^2}(1-t=1)$ because unit reflectivity implies zero transmissivity. In this case, the BS attack becomes equivalent to the IR attack, with Eve replacing the entire intercepted data sample by a new signal sent to Bob. 
This comparison among different supervised classifiers suggests that, in the considered regime, the discrimination performance is essentially classifier-independent, meaning that the distinguishability of the eight photon-number distributions is mainly governed by their first two statistical moments rather than by the specific decision rule. Since no classifier provides a systematic advantage, in the following we adopt the kNN classifier as a representative choice.
Remarkably, Bob's accuracy is always high despite an interception is occurring.
This is because the light signal that Eve sends to Bob in place of the original one sent by Alice has the same first two moments of the distribution Eve measures. Although the signal measured by Bob is no longer the one sent by Alice, his discrimination strategy is only mildly affected by the attacks. To properly handle them, Bob has to devise an alternative strategy to detect and counteract eavesdropping. 
\subsection{The eavesdropping attacks}
\begin{figure*}[!t]
\centering
\hfill
\includegraphics[width=0.9\textwidth]{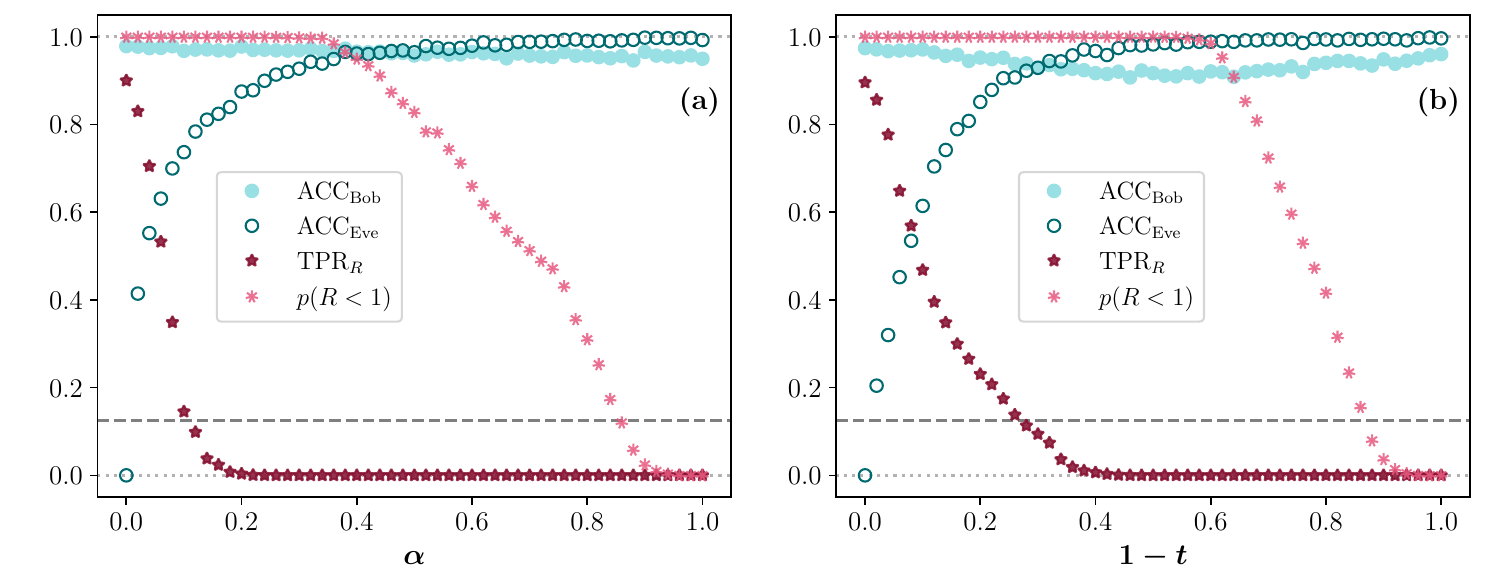}
\caption{Bob's (dots) and Eve's (open circles) discrimination accuracy ${\rm ACC_{\sigma^2}}$ as a function of (a) $\alpha$ in the IR attack and (b) $(1-t)$ in the BS attack. The dashed line indicates the accuracy of the random classifier, namely ${\rm ACC} = 1/8$. Security-related figures of merit based on the noise reduction factor $R$ are the probability that Bob measures nonclassical correlations $p(R<1)$ (asterisks) and the TPR$_R$ (stars). In both panels, the parameters are $\langle m_{\rm s,L} \rangle = 0.40$, $\langle m_{\rm s,H} \rangle = 0.45$, $\langle m_{\rm{Alice}} \rangle = 5$, $\Delta = 0.5$, and $d_{\rm sample} = 2\times 10^4$.}
\label{attacks}
\end{figure*}

We now detail the two eavesdropping strategies, IR and BS attacks, that Eve can carry out. We compare Bob's and Eve's discrimination capability in terms of ${\rm ACC}_{\sigma^2}$, and we devise proper countermeasures that Bob can implement against Eve to make the communication secure.
For both the IR attack [Fig.~\ref{attacks}(a)] and BS attack [Fig.~\ref{attacks}(b)], Bob's ${\rm ACC}_{\sigma^2}$ remains consistently high, exceeding $>0.92$ in the former case and $0.86$ in the latter, because Eve resends a light signal with the exact same first two moments of the original photon-number distribution. This means that accuracy alone is not sensitive enough to eavesdropping to ensure secure communication.
Crucially, Eve can discriminate the sent states with ${\rm ACC}_{\sigma^2} > 0.8$  even performing weak attacks, i.e., with $\alpha \sim 0.2$ and $(1-t) \sim 0.2$, thanks to the higher quantum efficiency of Eve's PNR detector ($\eta = 1$) relative to Bob's one ($\eta = 0.4$).
Bob cannot rely on the na\"ive criterion $p(R<1)$ (asterisks in Fig. \ref{attacks}) to ensure secure communication. Although the noise reduction factor is sensitive to correlation-breaking attacks, this criterion reveals the presence of an eavesdropper only for sufficiently strong IR and BS attacks, namely for $\alpha \gtrsim 0.35$ and $(1-t) \gtrsim 0.6$, respectively. Since Eve can already extract information through weaker attacks, her information-gain capability exceeds Bob's eavesdropping-detection capability, rendering the communication insecure.
\begin{figure*}[!t]
\centering
\hfill
\includegraphics[width=0.9\textwidth]{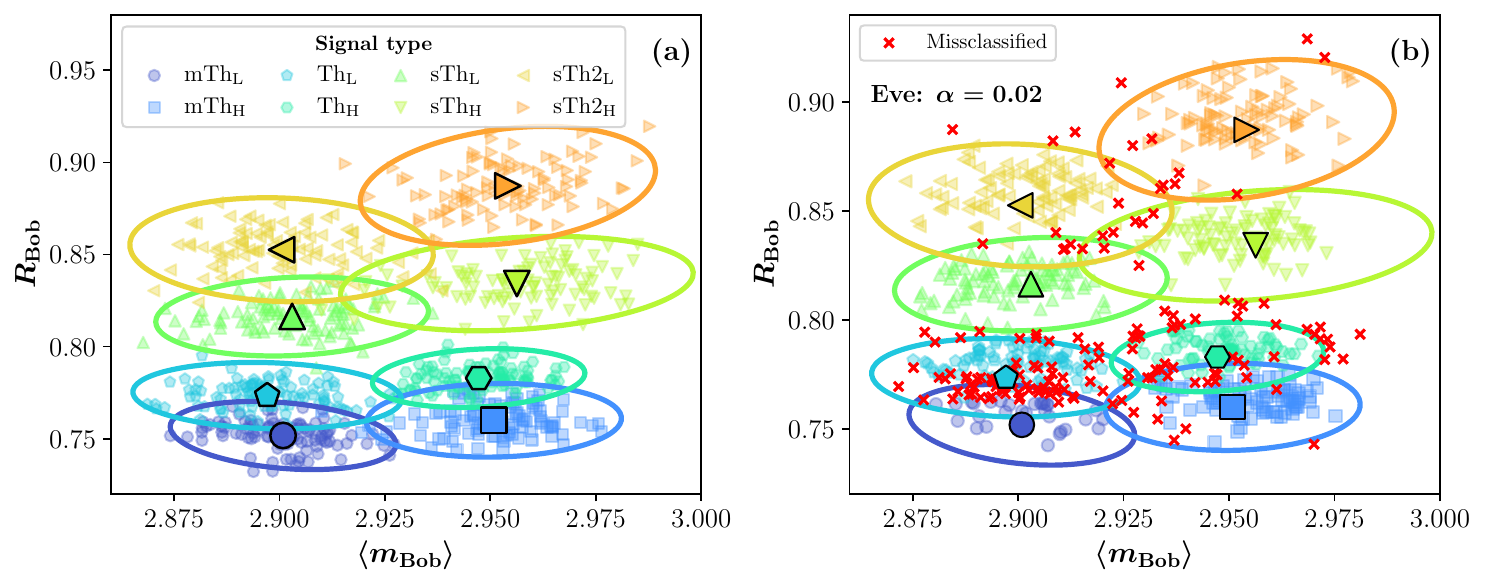}
\caption{(a) Noise reduction factor calculated by Bob as a function of the mean number of photons he detects in the case of different signals superimposed on the TWB: multi-mode pseudo-thermal states (mTh), single-mode pseudo-thermal states (Th), super-thermal states from a speckled-speckle field (sTh), and second-harmonic of single-mode pseudo-thermal states (sTh2). The subscripts L and H refer to low and high mean values of the signal states. The different symbols correspond to different realizations, obtained by the bootstrap procedure. Each large symbol associated with a signal denotes the reference values---mean value of $R$ and $m$---determined during the calibration stage to fix the center of the corresponding ellipse. Each ellipse encloses the 95$\%$ of data corresponding to a specific signal. (b) The same as in (a) in the case of an IR attack with $\alpha = 0.02$. The red crosses represent the signals that Bob classifies as potentially affected by eavesdropping.}
\label{ellipse}
\end{figure*}
These considerations motivate the introduction of a tailored security criterion that exploits the nonclassical photon-number correlations of the TWB to ensure secure communication.
The idea is to complement the discrimination of the transmitted symbol with a symbol-resolved security test based on the measured values of the noise reduction factor. Specifically, for each symbol we identify, under trusted conditions, a reference value for the noise reduction factor $R$ together with the corresponding acceptance criterion. A received symbol is regarded as secure if its measured $R$ satisfies the acceptance criterion; otherwise, it is discarded as indicative of eavesdropping. The proposed security criterion relies on the TPR$_R$, i.e., the probability that a secure symbol is correctly classified as secure, and is defined as follows.

Before the actual communication stage, Alice and Bob perform a calibration step in which each of the eight symbols is transmitted through the channel under trusted conditions. For each received sample, Bob estimates the mean number of detected photons and the noise reduction factor, thus obtaining a point in the two-dimensional space $\mathbf{x}= (\langle m_{\rm Bob}\rangle,R_{\rm Bob})$. For each symbol $j$, the corresponding reference cloud is characterized by its mean vector $\boldsymbol{\mu}_j$ and covariance matrix $\Sigma_j$. The acceptance region is then defined as the confidence ellipse
\begin{equation}
   (\mathbf{x}-\boldsymbol{\mu}_j)^T\Sigma_j^{-1}(\mathbf{x}-\boldsymbol{\mu}_j)
\leq
\chi^2_2(0.95), 
\end{equation}
where $\chi^2_2(0.95)$ is the 95th percentile of the chi-squared distribution with two degrees of freedom. 
For the photon-number distributions considered in the protocol, i.e. single-mode, multi-mode and super-thermal ones, and for the considered parameters ($\langle m_{\rm s,L} \rangle = 0.40$, $\langle m_{\rm s,H} \rangle = 0.45$, $\langle m_{\rm{Alice}} \rangle = 5.0$), the resulting reference clouds are shown in Fig.~\ref{ellipse}(a). The different points correspond to samples of $2\times10^4$ data obtained by applying the bootstrap procedure to the original datasets. The clouds are well described by ellipses, which are not aligned with the coordinate axes because the noise reduction factor depends on both the mean value and the variance of the detected photon-number distribution. This geometrical representation provides symbol-dependent thresholds that properly account for the covariance of each distribution.

During the communication stage, Alice communicates to Bob her measured photon number per pulse over an untrusted classical channel. Although this communication is potentially accessible to Eve, the shared data do not contain information about the encoded symbols. Bob uses these results to estimate $R$ for each data sample corresponding to a symbol.
Then, he discriminates the symbols using the local-moment strategy, and assesses the security of each symbol by testing the associated value of $R$: if the estimated value falls within the corresponding pre-calibrated acceptance ellipse, then the symbol is accepted; otherwise, it is rejected.
To illustrate this, Fig. \ref{ellipse}(b) shows the security validation of the symbols in the presence of a weak IR attack ($\alpha=0.02$). The high number of misclassified symbols (red crosses), corresponding to signals for which the acceptance criteria are not met, indicate that the TPR$_R$ is a sensitive indicator of the presence of an eavesdropper. 
Indeed, the TPR$_R$ (stars) in both panels of Fig.~\ref{attacks} exhibits a monotonically decreasing trend, dropping to zero at $\alpha = 0.2$ and $(1-t)=0.4$, respectively.
Furthermore, we point out that this procedure, based on pre-calibrated acceptance criteria, provides Bob with a security test that allows him to retain or discard received symbols without sacrificing useful data during communication. In contrast, without this prior calibration, potentially useful transmitted data would need to be sacrificed to determine, on a case-by-case basis, which symbol can be trusted.

\begin{figure*}[!t]
\centering
\hfill
\includegraphics[width=0.9\textwidth]{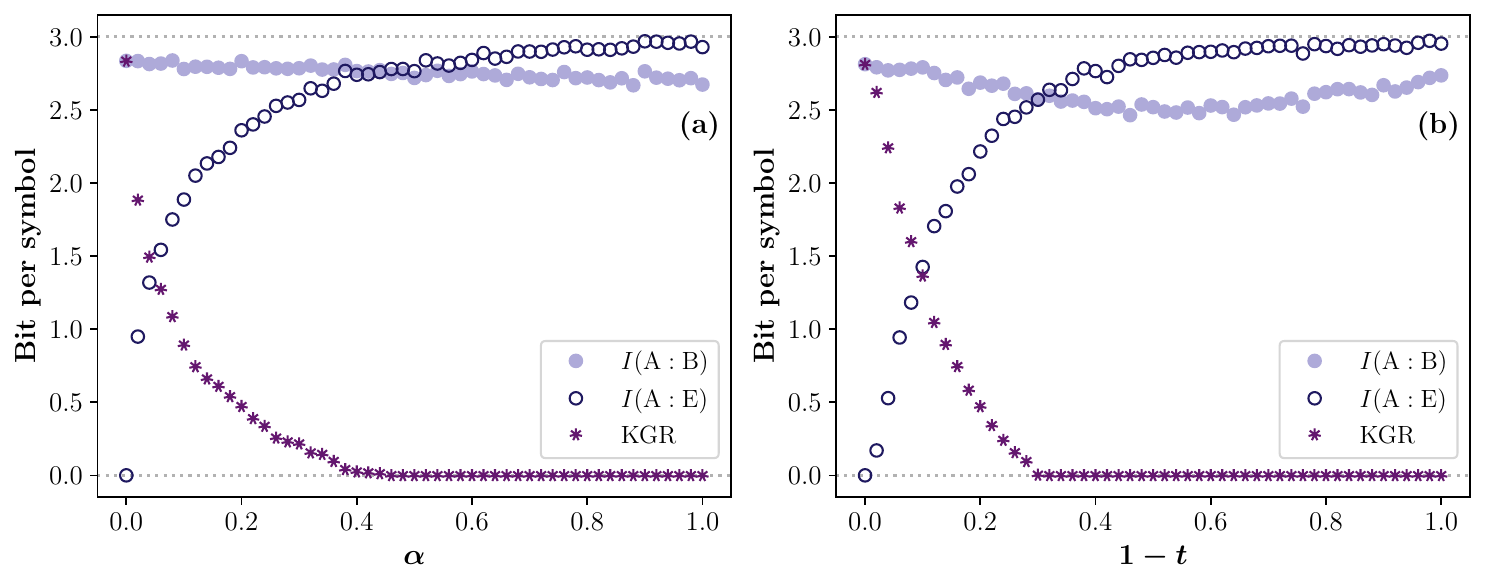}
\caption{MI between Alice and Bob, $I(A:B)$ (dots), MI between Alice and Eve, $I(A:E)$ (open circles), and KGR [Eq. \eqref{eq:kgr}] (asterisks; in bit per symbol) as a function of (a) $\alpha$ in the IR attack and (b) $(1-t)$ in the BS attack. In both panels, the parameters are $\langle m_{\rm s,L} \rangle = 0.40$, $\langle m_{\rm s,H} \rangle = 0.45$, $\langle m_{\rm{Alice}} \rangle = 5.0$, $\Delta = 0.5$ and $d_{\rm sample} = 2\times 10^4$.}
\label{MIetKGR}
\end{figure*}

Although the TPR$_R$ provides a convenient criterion to detect the presence of the eavesdropper, in principle it is not, by itself, sufficient to enable the generation of a secret key encoded in the eight symbols.
The key-generation capability of the protocol can be assessed through the KGR, which is defined as the difference of the MI between Alice and Bob $I(A:B)$ and that between Alice and Eve $I(A:E)$,
\begin{equation}
    \rm{KGR} = 
    \begin{cases}
        I(A:B) - I(A:E) & \text{if  $I(A:B) > I(A:E)$},\\
        0               & \text{otherwise.}
    \end{cases}
    \label{eq:kgr}
\end{equation}
Numerical results of the KGR are shown in {Fig.~\ref{MIetKGR}(a)} and {Fig.~\ref{MIetKGR}(b)} for IR and BS attacks, respectively. While the MI between Alice and Bob remains always close to the maximum, ideally equal to $\log_2 8 = 3$, the MI between Alice and Eve increases with increasing values of $\alpha$ and $(1-t)$, respectively.
Thus, there are regimes, i.e. $\alpha \lesssim 0.4$ and $(1-t) \lesssim0.3$, where KGR takes nonzero values, meaning that for some ranges a secure communication between Alice and Bob is possible despite the presence of an eavesdropper.
In addition, we point out that the values of KGR can be improved by performing privacy amplification \cite{bennett1988,bennett1995,cachin1997}. 
\begin{figure*}[!t]
\centering
\hfill
\includegraphics[width=0.9\textwidth]{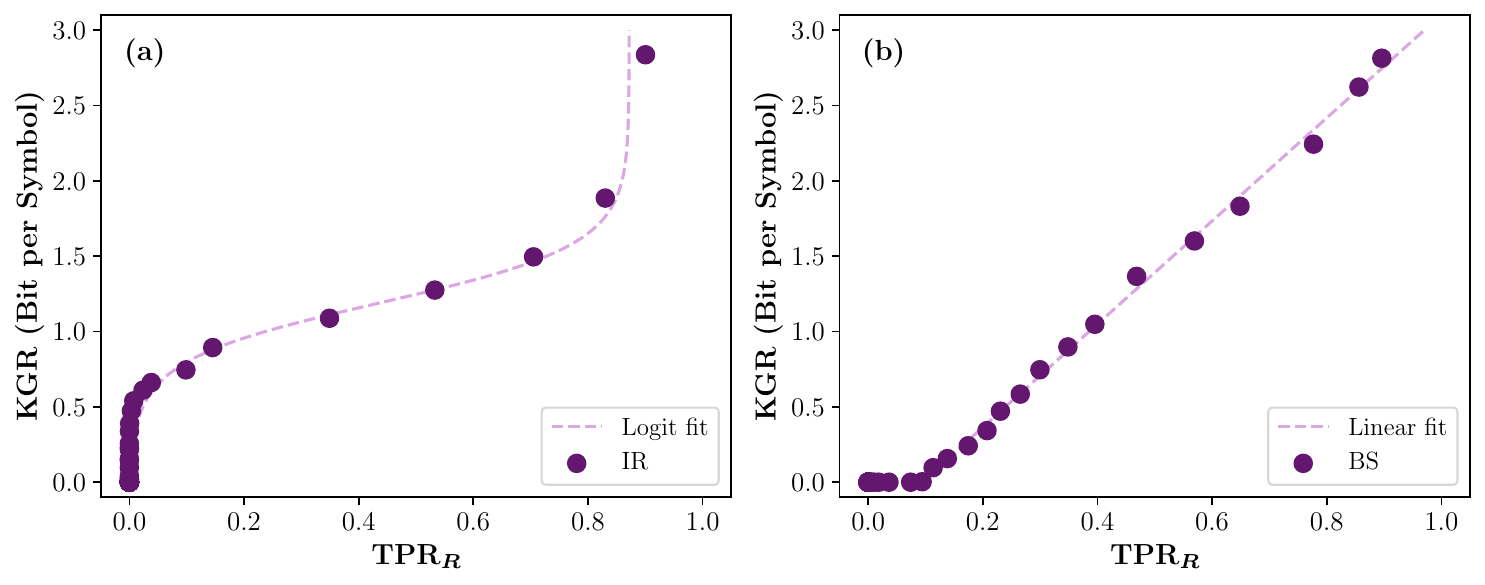}
\caption{KGR (in bit per symbol) as a function of the TPR$_R$ for (a) the IR attack and (b) the BS attack. In both panels the values of TPR$_{R}$ and KGR (dots) are those shown in Figs.~\ref{attacks} and \ref{MIetKGR}, while the dashed lines represent the fit with a logit function (panel (a)) and a linear function (panel (b)).} 
\label{KGRvsEC}
\end{figure*}
\section{Discussion}\label{sec_discussion}
By comparing Figs.~\ref{attacks} and \ref{MIetKGR}, we notice that for both the attacks the TPR$_R$ and the KGR exhibit analogous decreasing behavior as a function of $\alpha$ and $(1-t)$. This analogy could be ascribed to the complementarity of the amount of information gained by Bob and Eve, as all the signal lost by Bob is acquired by Eve. The relation between the two quantities is further investigated in Fig.~\ref{KGRvsEC}, where one quantity is plotted against the other, revealing a positive correlation between the two: in both cases, the KGR increases monotonically with the TPR$_R$.
More specifically, for IR attacks, the KGR as a function of the TPR$_R$ is well fitted by a logit curve,\footnote{The logit (logistic unit) function, $\operatorname{logit}(x) = f^{-1}(x) = \ln[x/(1-x)]$ with $x \in (0,1)$, is the inverse of the standard logistic function $f(x) = 1/(1+e^{-x})$.} meaning that the generation of a secret key remains possible even for very low TPR$_R$ values. For BS attacks, instead, larger TPR$_R$ values are needed to reach a non-zero KGR.
This proves that BS attacks are more detrimental to the key generation than IR attacks.

Higher KGR values can also be achieved by enlarging the alphabet, i.e., by using a larger number of symbols to encode information. In the present communication scheme, this can be reasonably accomplished by introducing additional mean values of the considered states or additional statistical distributions, such as coherent states or multi-mode super-thermal states.
The challenge is to properly account for the partial overlap among the states in Fig.~\ref{ellipse}: a certain degree of overlap can enhance security, whereas excessive overlap can compromise discrimination. This trade-off can be controlled by properly choosing the relevant parameters, such as the mean values of the signal states and the number of involved modes. In this regard, we point out that the present study adopts experimentally accessible parameter values that provide a suitable overlap among states.

We finally elaborate on the protocol's performance achievable in realistic scenarios, compared with already existing quantum communication schemes.
Standard quantum communication in optical fibers can achieve tens of Mbps over a distance of 10 km and up to a few Mbps over a distance of 100 km \cite{grunenfelder2023fast}. In our scheme, repetition rates of tens of MHz are achievable by suitably acting on the acquisition of PNR detectors output signal, thus being compatible with fiber communication. 
Quantum communication in free space, instead, can cover shorter distances, typically limited to tens of kms, due to the detrimental effects of the atmosphere, such as turbulence and scattering \cite{pirandola2021limits}. In this regard, we expect that mesoscopic optical states prove more robust against such effects, thus enabling long-distance free-space quantum communication.
Finally, inter-satellite quantum communication in space can reach higher rates and cover larger distances because of the absence of atmosphere. However, it requires detectors that can reliably operate under the harsh environmental conditions in space. In this regard,  the SiPMs \cite{chesi19} considered as PNR detectors in our protocol represent a suitable choice because they are immune to magnetic fields and do not require cryogenic temperature values \cite{ramilli,cassina}. 
Overall, these considerations highlight the potential of the mesoscopic intensity regime as a promising approach for future quantum communication protocols.

\section{Conclusions}\label{sec_conclusions}
In this work, we have numerically explored quantum communication in the mesoscopic intensity regime by proposing a protocol in which Alice encodes information in the statistical properties of different classical optical states, while Bob decodes it using PNR detectors, capable of counting the number of photons in each light pulse. The numerical simulations rely on experimentally accessible parameter regimes.
We assess the robustness of the protocol against two possible eavesdropping strategies, namely the IR and BS attacks, assuming that in both cases Eve replaces the subtracted light with a signal having the same mean value and variance. To ensure secure communication, we include a TWB state in the quantum channel through which the different signals propagate, and introduce a security criterion that exploits the nonclassical photon-number correlations of the TWB, quantified by the noise reduction factor, to reveal eavesdropping.
The effectiveness of the proposed quantum communication protocol is assessed in terms of its ability to generate a secret key shared by Alice and Bob. In this respect, the nonzero values of the KGR indicate that the protocol enables secure key generation against the considered eavesdropping attacks.

Overall, the numerical results obtained in this proof-of-principle study encourage the experimental realization of the proposed protocol.
To this aim, an experimental implementation would benefit from a plug-and-play source capable of switching reproducibly among the different signal states.
Regarding the detection system, SiPMs represent a suitable class of PNR detectors owing to their good dynamic range \cite{cassina} and the availability of models with efficiency up to 40$\%$, such as the S13360 series by Hamamatsu \cite{S13360}.
Moreover, data acquisition would benefit from the use of the digitizer we have recently exploited to characterize well-populated classical and quantum states of light (with mean values up to $\langle m \rangle = 30-40$) operating at high repetition rates (on the order of some MHz) \cite{arxiv2026}. 

\acknowledgments{
We thank Maristella Crotti (University of Insubria) for fruitful discussions. 
%
G.~C. acknowledges the Spanish State Research Agency through the María de Maeztu project CEX2021-001164-M and the COQUSY project PID2022-140506NB-C21 and -C22, all funded by MCIU/AEI/10.13039/501100011033. G.~C. further acknowledges the INFOLANET project PID2022-139409NB-I00 and the QuantERA QNet project PCI2024-153410, funded by MICIU/AEI/10.13039/501100011033 and by ERDF, EU.
L.~R. acknowledges support from University of Pavia through the project 
``Termodinamica di precisione per sistemi aperti quantistici'',
funded within the ``Fondo Ricerca e Giovani 2024'' programme, and from INFN through the project ``BELL''.
}
\section*{Author Contributions}
\textbf{Gabriele Cenedese:} Data curation (lead); Formal analysis (lead); Methodology (equal); Writing – Original draft (equal); Review and editing (equal). \textbf{Alex Pozzoli:} Data curation (equal); Methodology (equal); Writing – Original draft (equal); Review and editing (equal). \textbf{Luca Razzoli:} Methodology (equal); Writing – Original draft (equal); Review and editing (equal). \textbf{Alessia Allevi:} Conceptualization (lead); Methodology (equal); Writing – Original draft (equal); Review and editing (equal).

\section*{Data Availability Statement}
The data that support the findings of this study are available from the corresponding author upon reasonable request.

\bibliography{bibliography.bib}

@book{cariolaro,
  title={Quantum communications},
  author={Cariolaro, Gianfranco},
  volume={2},
  year={2015},
  publisher={Springer}
}

@article{flamini,
  title={Photonic quantum information processing: a review},
  author={Flamini, Fulvio and Spagnolo, Nicolo and Sciarrino, Fabio},
  journal={Reports on Progress in Physics},
  volume={82},
  number={1},
  pages={016001},
  year={2019},
  publisher={IOP Publishing},
  doi={10.1088/1361-6633/aad5b2}
}

@article{diamanti,
  title={Practical challenges in quantum key distribution},
  author={Diamanti, Eleni and Lo, Hoi-Kwong and Qi, Bing and Yuan, Zhiliang},
  journal={npj Quantum Information},
  volume={2},
  number={1},
  pages={16025},
  year={2016},
  publisher={Nature Publishing Group},
  doi={10.1038/npjqi.2016.25}
}

@article{dimario2018robust,
  title={Robust measurement for the discrimination of binary coherent states},
  author={DiMario, Matthew T and Becerra, Francisco E},
  journal={Physical Review Letters},
  volume={121},
  number={2},
  pages={023603},
  year={2018},
  publisher={APS},
  doi={10.1103/PhysRevLett.121.023603}
}

@article{dimario2,
  title={Optimized communication strategies with binary coherent states over phase noise channels},
  author={DiMario, MT and Kunz, L and Banaszek, K and Becerra, FE},
  journal={npj Quantum Information},
  volume={5},
  number={1},
  pages={65},
  year={2019},
  publisher={Nature Publishing Group UK London},
  doi={10.1038/s41534-019-0177-4}
}

@inproceedings{bennett1984quantum,
  title={Quantum cryptography: Public key distribution and con tos5},
  author={Bennett, Charles H and Brassard, Gilles},
  booktitle={Proceedings of the international conference on computers, systems and signal processing},
  pages={175--179},
  year={1984}
}

@article{bennett1992experimental,
  title={Experimental quantum cryptography},
  author={Bennett, Charles H and Bessette, Fran{\c{c}}ois and Brassard, Gilles and Salvail, Louis and Smolin, John},
  journal={Journal of cryptology},
  volume={5},
  number={1},
  pages={3--28},
  year={1992},
  publisher={Springer}
}

@article{ekert1991,
  title={Quantum cryptography based on Bell’s theorem},
  author={Ekert, Artur K},
  journal={Physical Review Letters},
  volume={67},
  number={6},
  pages={661},
  year={1991},
  publisher={APS}
}

@article{oe24,
  title={Assessing a binary quantum channel exploiting a silicon photomultiplier based hybrid receiver},
  author={Sanvito, Alberto and Cassina, Silvia and Lamperti, Marco and Notarnicola, Michele N and Olivares, Stefano and Allevi, Alessia},
  journal={Optics Express},
  volume={32},
  number={22},
  pages={39846--39859},
  year={2024},
  publisher={Optica Publishing Group},
  doi={10.1364/OE.534910}
}

@book{benenti,
  title={Principles of quantum computation and information: a comprehensive textbook},
  author={Benenti, Giuliano and Casati, Giulio and Rossini, Davide and Strini, Giuliano},
  year={2019},
  publisher={World Scientific}
}

@article{oe21,
  title={Effect of noisy channels on the transmission of mesoscopic twin-beam states},
  author={Allevi, Alessia and Bondani, Maria},
  journal={Optics Express},
  volume={29},
  number={21},
  pages={32842--32852},
  year={2021},
  publisher={Optical Society of America},
  doi={10.1364/OE.436079}
}

@article{scirep22,
  title={Novel scheme for secure data transmission based on mesoscopic twin beams and photon-number-resolving detectors},
  author={Allevi, Alessia and Bondani, Maria},
  journal={Scientific Reports},
  volume={12},
  number={1},
  pages={15621},
  year={2022},
  publisher={Nature Publishing Group UK London},
  doi={10.1038/s41598-022-19503-y}
}

@article{usenko,
  title={Continuous-variable quantum communication},
  author={Usenko, Vladyslav C and Ac{\'\i}n, Antonio and All{\'e}aume, Romain and Andersen, Ulrik L and Diamanti, Eleni and Gehring, Tobias and Hajomer, Adnan AE and Kanitschar, Florian and Pacher, Christoph and Pirandola, Stefano and others},
  journal={arXiv preprint arXiv:2501.12801},
  year={2025},
  doi={10.48550/arXiv.2501.12801}
}

@article{agliati,
  title={Quantum and classical correlations of intense beams of light investigated via joint photodetection},
  author={Agliati, Andrea and Bondani, Maria and Andreoni, Alessandra and De Cillis, Giovanni and Paris, Matteo GA},
  journal={Journal of Optics B: Quantum and Semiclassical Optics},
  volume={7},
  number={12},
  pages={S652--S663},
  year={2005},
  doi={10.1088/1464-4266/7/12/031}
}

@article{ramilli,
  title={Photon-number statistics with silicon photomultipliers},
  author={Ramilli, Marco and Allevi, Alessia and Chmill, Valery and Bondani, Maria and Caccia, Massimo and Andreoni, Alessandra},
  journal={Journal of the Optical Society of America B},
  volume={27},
  number={5},
  pages={852--862},
  year={2010},
  publisher={Optical Society of America},
  doi={10.1364/JOSAB.27.000852}
}

@book{mandel,
  title={Optical coherence and quantum optics},
  author={Mandel, Leonard and Wolf, Emil and Shapiro, Jeffrey H},
  year={1996},
  publisher={American Institute of Physics}
}

@article{arimondo,
  title={Nonlinear and quantum optical properties and applications of intense twin-beams},
  author={Allevi, Alessia and Bondani, Maria},
  journal={Advances In Atomic, Molecular, and Optical Physics},
  volume={66},
  pages={49--110},
  year={2017},
  publisher={Elsevier},
  doi={10.1016/bs.aamop.2017.02.001}
}

@article{pla22,
  title={Multi-mode twin-beam states in the mesoscopic intensity domain},
  author={Allevi, Alessia and Bondani, Maria},
  journal={Physics Letters A},
  volume={423},
  pages={127828},
  year={2022},
  publisher={Elsevier},
  doi={10.1016/j.physleta.2021.127828}
}

@article{epl10,
  title={Conditional measurements on multimode pairwise entangled states from spontaneous parametric downconversion},
  author={Allevi, Alessia and Andreoni, Alessandra and Beduini, Federica A and Bondani, Maria and Genoni, Marco G and Olivares, Stefano and Paris, Matteo GA},
  journal={EPL (Europhysics Letters)},
  volume={92},
  number={2},
  pages={20007},
  year={2010},
  doi={10.1209/0295-5075/92/20007}
}

@article{silberhorn2016,
  title={Single-mode parametric-down-conversion states with 50 photons as a source for mesoscopic quantum optics},
  author={Harder, Georg and Bartley, Tim J and Lita, Adriana E and Nam, Sae Woo and Gerrits, Thomas and Silberhorn, Christine},
  journal={Physical Review Letters},
  volume={116},
  number={14},
  pages={143601},
  year={2016},
  publisher={APS},
  doi={10.1103/PhysRevLett.116.143601}
}

@article{pla23,
  title={Thermal and superthermal noise signals as resources for underwater quantum communication},
  author={Allevi, Alessia and Bondani, Maria},
  journal={Physics Letters A},
  volume={492},
  pages={129207},
  year={2023},
  publisher={Elsevier},
  doi={10.1016/j.physleta.2023.129207}
}

@article{cassina,
  title={Exploiting the wide dynamic range of silicon photomultipliers for quantum optics applications},
  author={Cassina, Silvia and Allevi, Alessia and Mascagna, Valerio and Prest, Michela and Vallazza, Erik and Bondani, Maria},
  journal={EPJ Quantum Technology},
  volume={8},
  number={1},
  pages={4},
  year={2021},
  publisher={Springer Berlin Heidelberg},
  doi={10.1140/epjqt/s40507-021-00093-z}
}

@incollection{efron,
  title={Bootstrap methods: another look at the jackknife},
  author={Efron, Bradley},
  booktitle={Breakthroughs in statistics: Methodology and distribution},
  pages={569--593},
  year={1992},
  publisher={Springer},
  url={https://www.jstor.org/stable/2958830}
}

@article{becerra,
  title={Implementation of generalized quantum measurements for unambiguous discrimination of multiple non-orthogonal coherent states},
  author={Becerra, Francisco E and Fan, Jingyun and Migdall, Alan},
  journal={Nature communications},
  volume={4},
  number={1},
  pages={2028},
  year={2013},
  publisher={Nature Publishing Group UK London},
  doi={10.1038/ncomms3028}
}

@article{gisin,
  title={Quantum cryptography: an overview of quantum key distribution},
  author={Rusca, Davide and Gisin, Nicolas},
  journal={arXiv preprint arXiv:2411.04044},
  year={2024}
}

@article{cozzolino,
author = {Cozzolino, Daniele and Da Lio, Beatrice and Bacco, Davide and Oxenløwe, Leif Katsuo},
title = {High-Dimensional Quantum Communication: Benefits, Progress, and Future Challenges},
journal = {Advanced Quantum Technologies},
volume = {2},
number = {12},
pages = {1900038},
doi = {https://doi.org/10.1002/qute.201900038},
url = {https://advanced.onlinelibrary.wiley.com/doi/abs/10.1002/qute.201900038},
eprint = {https://advanced.onlinelibrary.wiley.com/doi/pdf/10.1002/qute.201900038},
year = {2019}
}

@article{boaron,
  title={High-speed integrated QKD system},
  author={Sax, Rebecka and Boaron, Alberto and Boso, Gianluca and Atzeni, Simone and Crespi, Andrea and Gr{\"u}nenfelder, Fadri and Rusca, Davide and Al-Saadi, Aws and Bronzi, Danilo and Kupijai, Sebastian and others},
  journal={Photonics Research},
  volume={11},
  number={6},
  pages={1007--1014},
  year={2023},
  publisher={Chinese Laser Press and Optica Publishing Group}
}

@article{chesi19,
author = {Giovanni Chesi and Luca Malinverno and Alessia Allevi and Romualdo Santoro and Massimo Caccia and Maria Bondani},
journal = {Opt. Lett.},
number = {6},
pages = {1371--1374},
publisher = {Optica Publishing Group},
title = {Measuring nonclassicality with silicon photomultipliers},
volume = {44},
month = {Mar},
year = {2019},
url = {https://opg.optica.org/ol/abstract.cfm?URI=ol-44-6-1371},
doi = {10.1364/OL.44.001371},
}

@article{endo,
  title={Optically sampled superconducting-nanostrip photon-number resolving detector for non-classical quantum state generation},
  author={Endo, Mamoru and Takahashi, Kazuma and Nomura, Takefumi and Sonoyama, Tatsuki and Miki, Shigehito and Yabuno, Masahiro and Terai, Hirotaka and Kashiwazaki, Takahiro and Inoue, Asuka and Umeki, Takeshi and others},
  journal={Optics Express},
  volume={33},
  number={15},
  pages={32545--32559},
  year={2025},
  publisher={Optica Publishing Group}
}

@article{calsamiglia,
  title = {Conditional beam-splitting attack on quantum key distribution},
  author = {Calsamiglia, John and Barnett, Stephen M. and L\"utkenhaus, Norbert},
  journal = {Phys. Rev. A},
  volume = {65},
  issue = {1},
  pages = {012312},
  numpages = {12},
  year = {2001},
  month = {Dec},
  publisher = {American Physical Society},
  doi = {10.1103/PhysRevA.65.012312},
  url = {https://link.aps.org/doi/10.1103/PhysRevA.65.012312}
}

@article{razzoli25,
doi = {10.1088/2058-9565/ae05c3},
url = {https://doi.org/10.1088/2058-9565/ae05c3},
year = {2025},
month = {sep},
publisher = {IOP Publishing},
volume = {10},
number = {4},
pages = {045036},
author = {Razzoli, Luca and Pozzoli, Alex and Allevi, Alessia},
title = {Hybrid discrimination strategy in quantum communication based on photon-number-resolving detectors and mesoscopic twin-beam states},
journal = {Quantum Science and Technology}
}

@article{osisanwo,
  title={Supervised machine learning algorithms: classification and comparison},
  author={Osisanwo, Folorunso Y and Akinsola, Joseph ET and Awodele, Oludele and Hinmikaiye, John O and Olakanmi, Oluwole and Akinjobi, Joseph and others},
  journal={International Journal of Computer Trends and Technology (IJCTT)},
  volume={48},
  number={3},
  pages={128--138},
  year={2017}
}

@article{kotsiantis,
  title={Machine learning: a review of classification and combining techniques},
  author={Kotsiantis, Sotiris B and Zaharakis, Ioannis D and Pintelas, Panayiotis E},
  journal={Artificial Intelligence Review},
  volume={26},
  number={3},
  pages={159--190},
  year={2006},
  publisher={Springer}
}

@article{bianciardi,
AUTHOR = {Bianciardi, Camilla and Allevi, Alessia and Bondani, Maria},
TITLE = {Experimental Validation of the Statistical Properties of Speckled-Speckle Fields in the Mesoscopic Intensity Regime},
JOURNAL = {Applied Sciences},
VOLUME = {13},
YEAR = {2023},
NUMBER = {7},
ARTICLE-NUMBER = {4490},
URL = {https://www.mdpi.com/2076-3417/13/7/4490},
ISSN = {2076-3417},
DOI = {10.3390/app13074490}
}

@article{ol15,
author = {Alessia Allevi and Maria Bondani},
journal = {Opt. Lett.},
number = {13},
pages = {3089--3092},
publisher = {Optica Publishing Group},
title = {Direct detection of super-thermal photon-number statistics in second-harmonic generation},
volume = {40},
month = {Jul},
year = {2015},
url = {https://opg.optica.org/ol/abstract.cfm?URI=ol-40-13-3089},
doi = {10.1364/OL.40.003089},
}

@article{izumi,
  title = {Displacement receiver for phase-shift-keyed coherent states},
  author = {Izumi, Shuro and Takeoka, Masahiro and Fujiwara, Mikio and Pozza, Nicola Dalla and Assalini, Antonio and Ema, Kazuhiro and Sasaki, Masahide},
  journal = {Phys. Rev. A},
  volume = {86},
  issue = {4},
  pages = {042328},
  numpages = {11},
  year = {2012},
  month = {Oct},
  publisher = {American Physical Society},
  doi = {10.1103/PhysRevA.86.042328},
  url = {https://link.aps.org/doi/10.1103/PhysRevA.86.042328}
}

@article{muller,
doi = {10.1088/1367-2630/17/3/032003},
url = {https://doi.org/10.1088/1367-2630/17/3/032003},
year = {2015},
month = {mar},
publisher = {IOP Publishing},
volume = {17},
number = {3},
pages = {032003},
author = {Müller, Christian R and Marquardt, Christoph},
title = {A robust quantum receiver for phase shift keyed signals},
journal = {New Journal of Physics}
}

@article{becerra1,
  title={Experimental demonstration of a receiver beating the standard quantum limit for multiple nonorthogonal state discrimination},
  author={Becerra, Francisco E and Fan, Jingyun and Baumgartner, G and Goldhar, JTKJ and Kosloski, JT and Migdall, A},
  journal={Nature Photonics},
  volume={7},
  number={2},
  pages={147--152},
  year={2013},
  publisher={Nature Publishing Group UK London}
}

@online{S13360,
  author = {{Hamamatsu Photonics}},
  title  = {MPPC (Multi-Pixel Photon Counter) S13360 series},
  url    = {https://www.hamamatsu.com/content/dam/hamamatsu-photonics/sites/documents/99_SALES_LIBRARY/ssd/s13360_series_kapd1052e.pdf},
  year   = {2025},
  urldate = {2026-04-30}
}

@article{arxiv2026,
  title={Developing a photon-number-resolving detection chain for quantum communication protocols involving mesoscopic states of light},
  author={Pozzoli, Alex and Carsi, Stefano and Abba, Andrea and Allevi, Alessia},
  journal={arXiv preprint arXiv:2605.19980},
  volume={},
  number={},
  year={2026},
  publisher={}
}

@article{machulka2014,
  title={Spatial properties of twin-beam correlations at low-to high-intensity transition},
  author={Machulka, Radek and Haderka, Ond{\v{r}}ej and Pe{\v{r}}ina Jr, Jan and Lamperti, Marco and Allevi, Alessia and Bondani, Maria},
  journal={Optics Express},
  volume={22},
  number={11},
  pages={13374--13379},
  year={2014},
  publisher={Optical Society of America}
}

@article{notarnicola2023,
  title={Hybrid near-optimum binary receiver with realistic photon-number-resolving detectors},
  author={Notarnicola, Michele N and Paris, Matteo GA and Olivares, Stefano},
  journal={Journal of the Optical Society of America B},
  volume={40},
  number={4},
  pages={705--714},
  year={2023},
  publisher={Optica Publishing Group}
}

@article{notarnicola2025,
  title={Employing weak-field homodyne detection for optical communications},
  author={Notarnicola, Michele N and Olivares, Stefano},
  journal={IEEE Journal on Selected Areas in Communications},
  year={2025},
  publisher={IEEE}
}

@book{goodman2007,
  title={Speckle phenomena in optics: theory and applications},
  author={Goodman, Joseph W},
  year={2007},
  publisher={Roberts and company Publishers}
}

@article{cattaneo2018,
  title={Hybrid quantum key distribution using coherent states and photon-number-resolving detectors},
  author={Cattaneo, Marco and Paris, Matteo GA and Olivares, Stefano},
  journal={Physical Review A},
  volume={98},
  number={1},
  pages={012333},
  year={2018},
  publisher={APS}
}

@article{bennett1988,
  title={Privacy amplification by public discussion},
  author={Bennett, Charles H and Brassard, Gilles and Robert, Jean-Marc},
  journal={SIAM journal on Computing},
  volume={17},
  number={2},
  pages={210--229},
  year={1988},
  publisher={SIAM}
}

@article{bennett1995,
  title={Generalized privacy amplification},
  author={Bennett, Charles H and Brassard, Gilles and Cr{\'e}peau, Claude and Maurer, Ueli M},
  journal={IEEE Transactions on Information theory},
  volume={41},
  number={6},
  pages={1915--1923},
  year={1995},
  publisher={IEEE}
}

@article{cachin1997,
  title={Linking information reconciliation and privacy amplification},
  author={Cachin, Christian and Maurer, Ueli M},
  journal={journal of Cryptology},
  volume={10},
  number={2},
  pages={97--110},
  year={1997},
  publisher={Springer}
}

@article{gaidash2016,
 title={Revealing beam-splitting attack in a quantum cryptography system with a photon-number-resolving detector},
 author={Gaidash, Andrei and Egorov, Vladimir and Gleim, Artur},
 journal={Journal of the Optical Society of America B},
 volume={33},
 number={7},
 pages={1451--1455},
 year={2016},
 publisher={Optical Society of America}
}

@article{scarani2009RevModPhys,
  title = {The security of practical quantum key distribution},
  author = {Scarani, Valerio and Bechmann-Pasquinucci, Helle and Cerf, Nicolas J. and Du\ifmmode \check{s}\else \v{s}\fi{}ek, Miloslav and L\"utkenhaus, Norbert and Peev, Momtchil},
  journal = {Rev. Mod. Phys.},
  volume = {81},
  issue = {3},
  pages = {1301--1350},
  numpages = {0},
  year = {2009},
  month = {Sep},
  publisher = {American Physical Society},
  doi = {10.1103/RevModPhys.81.1301},
  url = {https://link.aps.org/doi/10.1103/RevModPhys.81.1301}
}

@article{xu2020RevModPhys,
  title = {Secure quantum key distribution with realistic devices},
  author = {Xu, Feihu and Ma, Xiongfeng and Zhang, Qiang and Lo, Hoi-Kwong and Pan, Jian-Wei},
  journal = {Rev. Mod. Phys.},
  volume = {92},
  issue = {2},
  pages = {025002},
  numpages = {60},
  year = {2020},
  month = {May},
  publisher = {American Physical Society},
  doi = {10.1103/RevModPhys.92.025002},
  url = {https://link.aps.org/doi/10.1103/RevModPhys.92.025002}
}

@article{eisaman2011revsciinst,
    author = {Eisaman, M. D. and Fan, J. and Migdall, A. and Polyakov, S. V.},
    title = {Invited Review Article: Single-photon sources and detectors},
    journal = {Review of Scientific Instruments},
    volume = {82},
    number = {7},
    pages = {071101},
    year = {2011},
    month = {07},
    issn = {0034-6748},
    doi = {10.1063/1.3610677},
    url = {https://doi.org/10.1063/1.3610677},
}

@article{li2020enhanced,
  title={Enhanced photon communication through Bayesian estimation with an SNSPD array},
  author={Li, Xiang and Tan, Jingrou and Zheng, Kaimin and Zhang, Labao and Zhang, Lijian and He, Weiji and Huang, Pengwei and Li, Haochen and Zhang, Biao and Chen, Qi and others},
  journal={Photonics Research},
  volume={8},
  number={5},
  pages={637--641},
  year={2020},
  publisher={Chinese Laser Press and Optical Society of America}
}

@article{azuma2024heralded,
  title={Heralded single-photon source based on superpositions of squeezed states},
  author={Azuma, Hiroo and Munro, William J and Nemoto, Kae},
  journal={Physical Review A},
  volume={109},
  number={5},
  pages={053711},
  year={2024},
  publisher={APS}
}

@online{bipm,
  author = {{BIPM, IEC, IFCC, ILAC, ISO, IUPAC, IUPAP, and OIML}},
  title  = {Evaluation of Measurement Data—Supplement 1 to the “Guide to the Expression of Uncertainty in Measurement”—Propagation of Distributions Using a Monte Carlo Method (JCGM 101)},
  url    = {www.bipm.org/en/doi/10.59161/jcgm101-2008},
  year   = {2008},
  urldate = {2026-06-24}
}

@article{grunenfelder2023fast,
  title={Fast single-photon detectors and real-time key distillation enable high secret-key-rate quantum key distribution systems},
  author={Gr{\"u}nenfelder, Fadri and Boaron, Alberto and Resta, Giovanni V and Perrenoud, Matthieu and Rusca, Davide and Barreiro, Claudio and Houlmann, Rapha{\"e}l and Sax, Rebecka and Stasi, Lorenzo and El-Khoury, Sylvain and others},
  journal={Nature Photonics},
  volume={17},
  number={5},
  pages={422--426},
  year={2023},
  publisher={Nature Publishing Group UK London}
}

@article{pirandola2021limits,
  title={Limits and security of free-space quantum communications},
  author={Pirandola, Stefano},
  journal={Physical Review Research},
  volume={3},
  number={1},
  pages={013279},
  year={2021},
  publisher={APS}
}
\end{document}